\documentclass[onecolumn]{aastex61}
\usepackage{amsmath}

\received{March 19, 2018}
\revised{June 5, 2018}
\accepted{June 18, 2018}
\submitjournal{AJ}

\shorttitle{Probing the Hill Sphere of (486958) 2014 MU$_{69}$}
\shortauthors{Kammer et al.}

\begin{document}

\title{Probing the Hill Sphere of (486958) 2014 MU$_{69}$:\\HST FGS Observations during the July 17, 2017 Stellar Occultation}

\correspondingauthor{Joshua A. Kammer}
\email{jkammer@swri.edu}

\author{Joshua A. Kammer}
\affil{Southwest Research Institute \\
San Antonio, TX 78238, USA}

\author{Tracy M. Becker}
\affil{Southwest Research Institute \\
San Antonio, TX 78238, USA}

\author{Kurt D. Retherford}
\affil{Southwest Research Institute \\
San Antonio, TX 78238, USA}
\affil{University of Texas at San Antonio \\
San Antonio, TX 78249, USA}

\author{S. Alan Stern}
\affil{Southwest Research Institute \\
Boulder, CO 80302, USA}

\author{Catherine B. Olkin}
\affil{Southwest Research Institute \\
Boulder, CO 80302, USA}

\author{Marc W. Buie}
\affil{Southwest Research Institute \\
Boulder, CO 80302, USA}

\author{John R. Spencer}
\affil{Southwest Research Institute \\
Boulder, CO 80302, USA}

\author{Amanda S. Bosh}
\affil{Massachusetts Institute of Technology \\
Cambridge, MA 02139, USA}

\author{Lawrence H. Wasserman}
\affil{Lowell Observatory \\
Flagstaff, AZ 86001, USA}

\begin{abstract}

We observed the July 17, 2017 stellar occultation of HD 168233 by the Kuiper Belt Object (486958) 2014 MU$_{69}$, the close flyby target of the extended New Horizons mission. Rather than capture a solid body occultation by the KBO itself, our program aimed to constrain the opacity of rings, moons, or other debris in the nearby environment. We used the Hubble Space Telescope Fine Guidance Sensors (HST FGS) instrument in TRANS~F583W mode to collect 40 Hz time resolution photometry of the stellar occultation star for two HST orbits during this observation. We present the results of reduction and calibration of the HST FGS photometry, and set upper limits on rings or other dust opacity within the Hill sphere of (486958) 2014 MU$_{69}$ at distances ranging from $\sim$20,000 km to $\sim$75,000 km from the main body.

\end{abstract}


\section{Introduction} \label{sec:intro}

The Kuiper belt object (486958) 2014 MU$_{69}$ (hereafter MU69) was discovered in June 2014 using the Hubble Space Telescope. It is part of the cold classical Kuiper belt, a structure that has been mostly undisturbed since the formation of the solar system. MU69 has a semi-major axis of 44.4 au and orbits the sun with a period of 295 years. Based on its apparent V-magnitude near 27, the diameter of the body is likely to be 20 to 40 km, depending on surface albedo. The recent results of \citet{buie2018} constrain its size further, as well as indicate it is likely a contact binary and may have at least one satellite. It is the current target of the NASA New~Horizons spacecraft, which is on its way towards a planned January 1, 2019 flyby of this distant object \citep{stern2018}. 

Pre-encounter observations of potential rings or dust in the environment of MU69 provide a critical role in avoiding spacecraft destruction given the 14.4~km/s New~Horizons flyby speed. Rings or other dust opacity structures have previously been detected during stellar occultations around two Centaurs (KBO escapees): 10199~Chariklo \citep{braga2014}, and 2060~Chiron \citep{ruprecht2013,ruprecht2015,ortiz2015}. In the case of Chariklo, these rings are located about 400~km from the center of a body that is roughly 250~km in diameter; for Chiron the rings (or possibly some other dust structure type) are about 320~km from the center of a body that is roughly 220~km in diameter. More recently, stellar occultation observations revealed that the dwarf planet Haumea also has a ring \citep{ortiz2017}. This large Trans-Neptunian object (TNO) has a volume-equivalent diameter of $\sim$1595$\pm$11~km, with the rings located approximately 2287 km from the center of the body.

MU69 is much smaller than these bodies, but its lower gravity could still maintain at least a transient amount of dust or debris in the system. Analyses of the more similarly-sized moons of Pluto, Nix (radius $\sim$44~km) and Hydra (radius $\sim$36~km), also showed that the low escape velocities of these objects can enable a significant amount of mass to escape given the micrometeorite bombardment rates expected in the Kuiper Belt \citep{stern2006,stern2009,poppe2011}. Additionally, stellar occultations of MU69 itself suggest that the object is likely a binary or contact binary \citep{buie2018}. Binary and multiple systems are predicted to make up at least 20\% of the cold, classical Kuiper Belt population \citep{stephens2006}. If MU69 does have an additional, widely-separated companion in its system, that object could serve as a potential source for an impact-induced, dusty ring far from the central body, perhaps in a similar way that the Phoebe ring extends radially to $\sim$18\% of Saturn's Hill sphere \citep{verbiscer2009}.

For this reason, HST GO/DD program 15003 was designed to make a deep probe of the Hill sphere region of MU69 during the July 17, 2017 stellar occultation. We discuss the background and planning of the observations in \S\ref{sec:observations} and outline the HST FGS data reduction process in \S\ref{sec:reduction}. We estimate upper limits on opacity and discuss these results in \S\ref{sec:limits} and \S\ref{sec:discussion}, respectively.

\section{HST FGS Observation Planning and Viewing Geometry} \label{sec:observations}

The Hubble Space Telescope Fine Guidance Sensors (HST FGS) are large field of view white light interferometers generally used for precision pointing of the telescope, but which also function as a science instrument capable of relative astrometry, detection of close binary systems, and 40-Hz relative photometry with milli-magnitude accuracy \citep{nelan2017}. This last feature makes it a powerful platform for observing stellar occultations by smaller bodies, which often occur over very short timescales ($\sim$seconds). HST FGS have previously been used to observe stellar occultations by solar system objects, as in the case of Triton \citep{elliot1998,elliot2000}.

Unlike for relatively larger bodies like Triton, the stellar occultation by MU69 cast a much smaller shadow, and therefore the odds of HST being able to observe the solid body passing in front of the star were exceedingly low. However, the event presented an opportunity to probe the nearby environment of MU69, in case significant opacity existed due to rings or some other debris source. We approximate the Hill sphere radius of MU69 as:

\begin{equation} 
r = a \, (1-e) \, \sqrt[3]{\frac{m}{3M_{\bigodot}}}
\end{equation}

\noindent with a semi-major axis $a$ of 44.4~au; an eccentricity $e$ of 0.04; and the mass of the sun, $M_{\bigodot}$, of $1.989 \times 10^{30}$ kg. We assumed a conservatively small diameter of 20~km and a mean density of 2~g/cm$^3$ to estimate that the mass, $m$, of MU69 would be on the order of $10^{16}$~kg. Therefore, the Hill sphere radius would be $\sim$75,000 km. This region casts a much larger shadow during the stellar occultation than the central body itself. For that reason, the observations acquired during this program were designed to cover two HST orbits overlapping the mid-time of the stellar occultation as seen from Earth on July~17,~2017.

\begin{figure*}[ht!]
\plotone{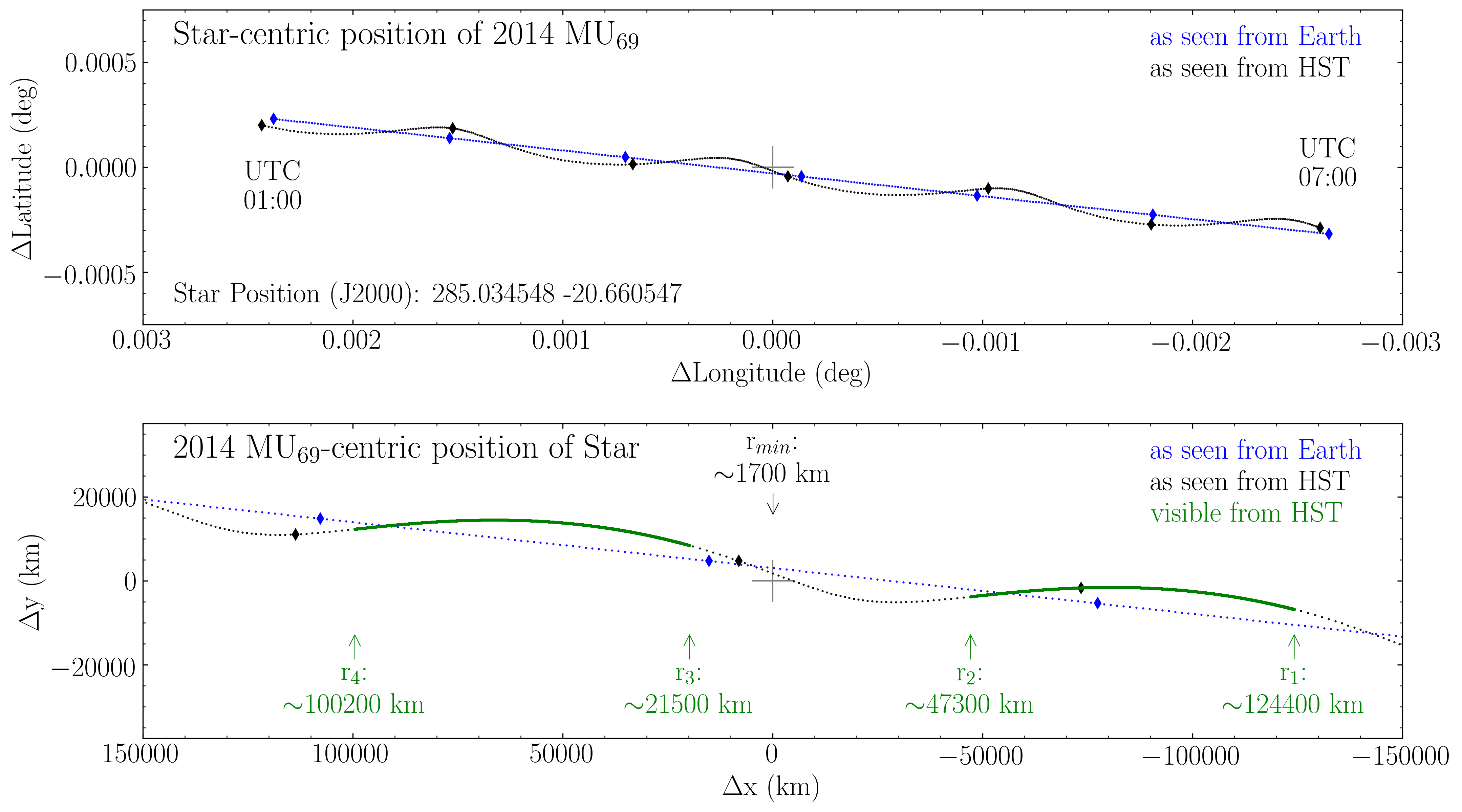}
\caption{Viewing geometry of the July 17, 2017 occultation based on the most recent MU69 orbital reconstruction. Top panel: Star-centered view of the apparent sky path of MU69 during the occultation. Axes are in units of ecliptic latitude and longitude relative to the position of the star. Points are shown for each hour from 1:00 to 7:00 UTC for the position of MU69 as seen from the center of Earth (blue) and as seen from the vantage point of HST (black). Bottom panel: MU69-centered view of the tangent radius of the star position over time, in units of km from the body. Apparent sky path of the star is shown for viewing from the center of Earth (blue) and from HST (black). Due to its orbit around Earth, the occultation star was only in view from HST during the locations shown in green, which correspond to the two orbits of data acquired in this program. The tangent radius (the projected distance from MU69 center to the line~of~sight of the star) at the beginning and end of each orbit are indicated (r$_1$,~r$_2$,~r$_3$,~r$_4$), as well as the minimum tangent radius that occurred while the view was occulted by Earth (r$_{min}$).\label{fig:geometry}}
\end{figure*}

Due to the orbital geometry of HST, it was inevitable that during part of the stellar occultation window the Earth itself would occult the view of MU69 and preclude the measurement of useful data. Unfortunately, this occurred near the mid-point of the occultation when the projected distance between the star and MU69 was at a minimum, as shown in Figure \ref{fig:geometry}. However, the observation by HST still probed a region of the MU69 Hill sphere at radii down to $\sim$20,000~km. Uncertainty in the sky-projected position of MU69 based on the ephemeris used in this work was approximately 20~km at the time of the observation.

\section{Data Reduction} \label{sec:reduction}

\begin{figure*}[ht!]
\plotone{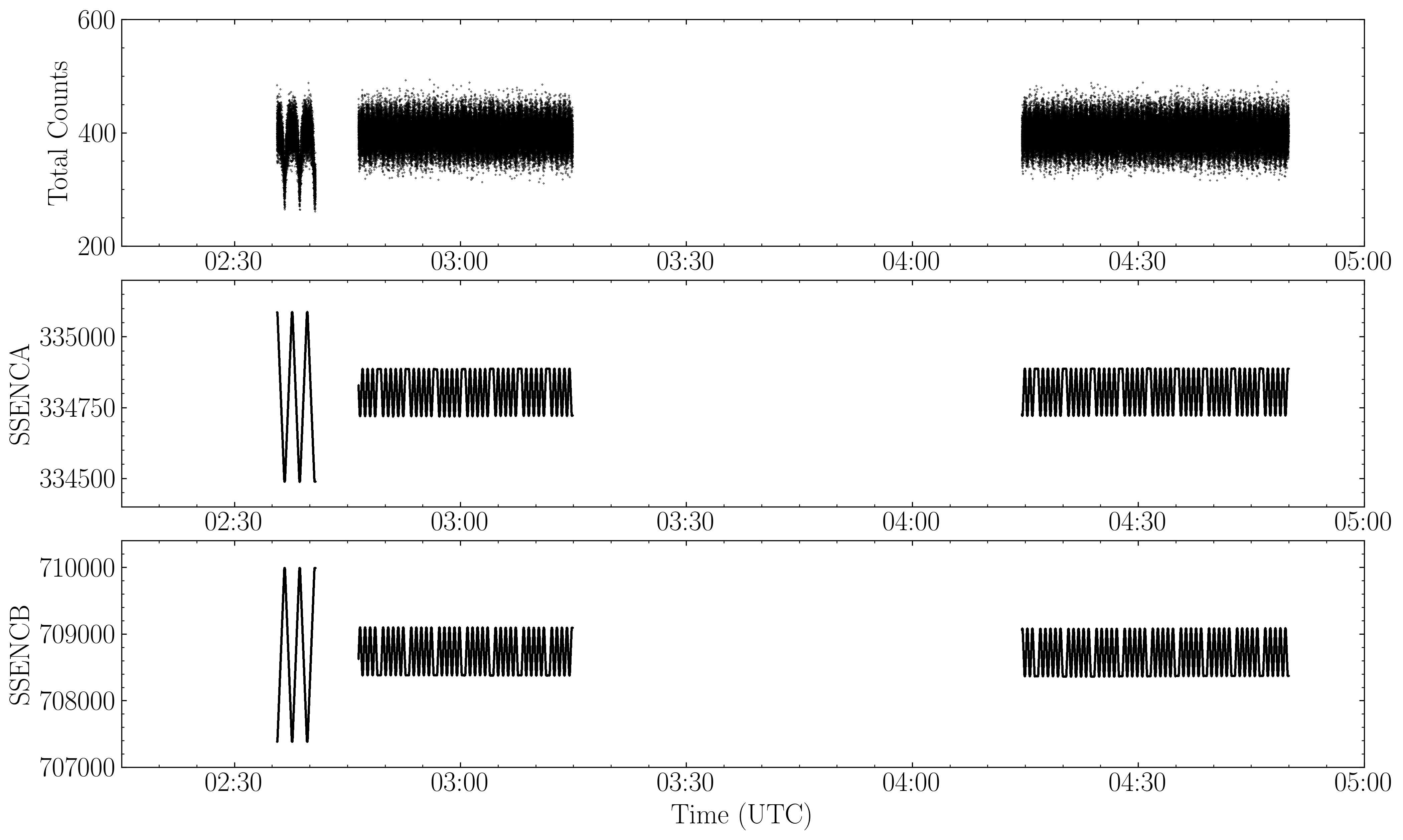}
\caption{Observed HST FGS counts during the occultation window. Top panel: The total counts from all four PMTs. Middle panel: The time-varying encoded position values of Star Selector A. Bottom panel: The time-varying encoded position values of Star Selector B. These values correspond to changes in the total count rate as the FGS instrument performs multiple scans across the stellar occultation target star.\label{fig:rawdata}}
\end{figure*}

The measurements acquired by HST FGS are initially reported in raw data counts at 40-Hz time resolution for each of the 4 photo-multiplier tubes (PMTs), as well as the encoded Star Selector A and B positions (SSENCA and SSENCB). Additionally, the data files contain flags updated every 150~msec that store the ongoing status of the instrument, including whether the target star has been detected and tracking has been engaged. For purposes of photometric analysis, we filtered the data using these flags to avoid instrument slews and only examined times when the instrument was tracking the target star in FineLock mode \citep{nelan2002}. We then summed the counts from all 4 PMTs into the total measured counts shown in Figure \ref{fig:rawdata}. The data from the first orbit consists of two segments: a wider $\sim$5~minute scan in order to rule out the presence of a binary companion near the target star; and a more focused $\sim$30~minute scan to monitor stellar photometry. This scan mode was then continued in the second HST orbit for another $\sim$35~minutes of data.

\begin{figure*}[ht!]
\centering
\includegraphics[width=0.75\textwidth,height=\textheight,keepaspectratio]{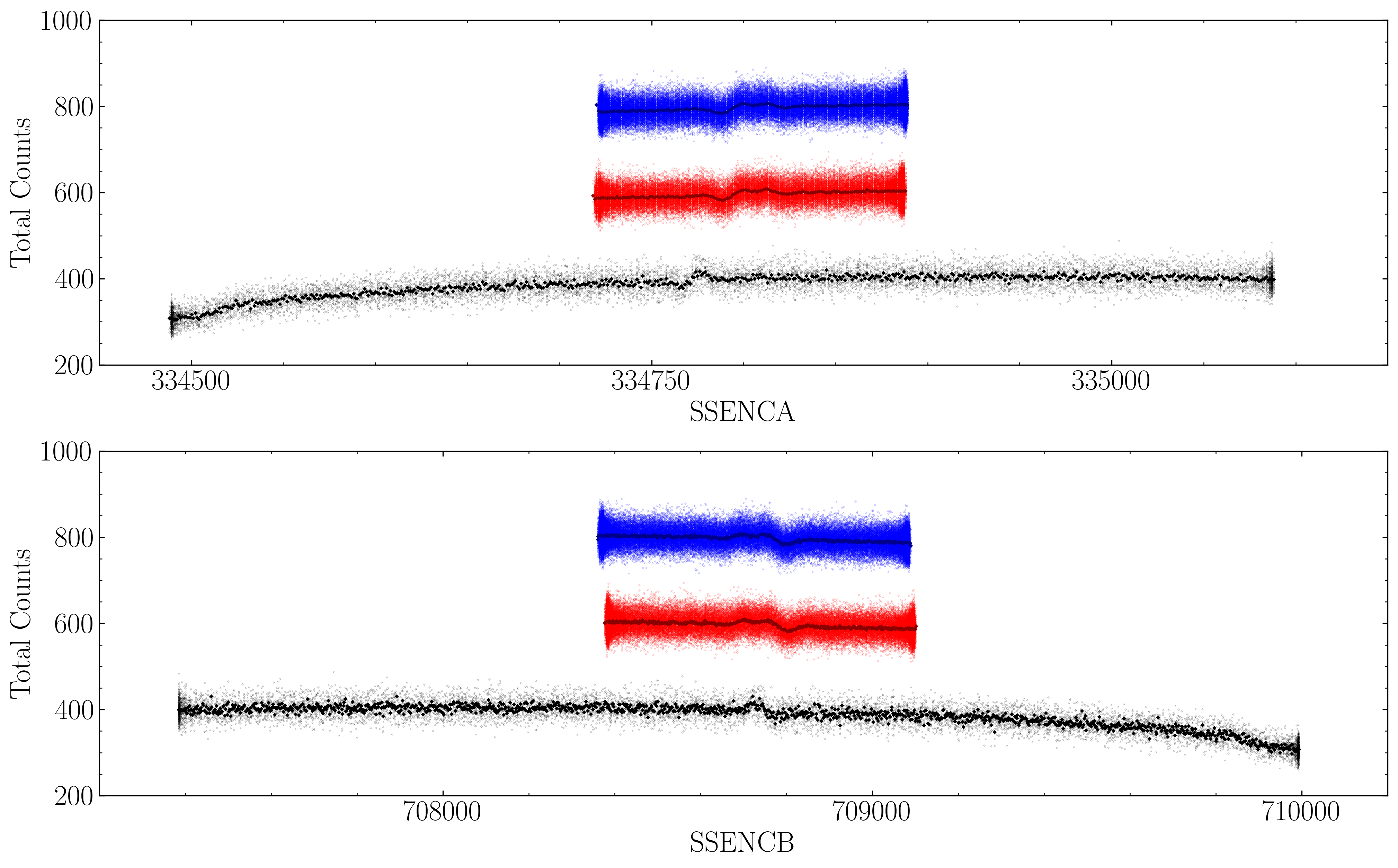}
\caption{Relationship between SSENCA, SSENCB, and total measured counts ($C_{obs}$). Each scan of the FGS instrument in TRANS mode results in a slightly different count rate that correlates with the encoded SSENCA and SSENCB values. An estimate of this functional relationship can be derived from the mean number of counts at a given value of SSENCA and SSENCB. The full set of 40-Hz data points are shown for: segment 1 of orbit 1 (gray), segment 2 of orbit 1 (red, offset by +200 counts), and orbit 2 (blue, offset by +400 counts). Similarly, the average value of $C_{obs}$ as a function of SSENCA and SSENCB is shown for: segment 1 of orbit 1 (black), segment 2 of orbit 1 (dark red, offset by +200 counts), and orbit 2 (dark blue, offset by +400 counts).}\label{fig:SSENC}
\end{figure*}

The measured FGS count rate is sensitive to the instantaneous pointing of the instrument, and this effect will dominate over relatively small signals of opacity along the line of sight. However, the encoded Star Selector A and B positions allow for an empirical fit to determine the relationship between these pointing positions and the total measured count rate. This is shown in Figure \ref{fig:SSENC}, where the measurement of total counts is seen to be a consistent function of the encoded SSENCA and SSENCB values.

We constructed a simple forward model prediction of total counts based on the known values of SSENCA and SSENCB during each segment of the HST observation. At each time step the expected variation in observed counts due solely to instrument effects, $C_{inst}(t)$, is derived from a weighted mean of the counts measured at all other times when the sensor was in the same scan position, as described in Equations \ref{eq:SSENC1} and \ref{eq:SSENC2}.

\begin{equation}\label{eq:SSENC1}
w_i(t) = \begin{cases}
1 &\mbox{if } \begin{aligned} t&\neq t_i\\ \text{SSENCA}(t)&=\text{SSENCA}(t_i)\\ \text{SSENCB}(t)&=\text{SSENCB}(t_i) \end{aligned} \\
0 &\mbox{otherwise} \\
\end{cases}
\end{equation}

\begin{equation}\label{eq:SSENC2}
C_{inst}(t) = \frac{\sum_{i=1}^n w_i(t)\: C_{obs}(t_i)}{\sum_{i=1}^n w_i(t)}
\end{equation}

\begin{figure*}[ht!]
\centering
\includegraphics[width=0.75\textwidth,height=\textheight,keepaspectratio]{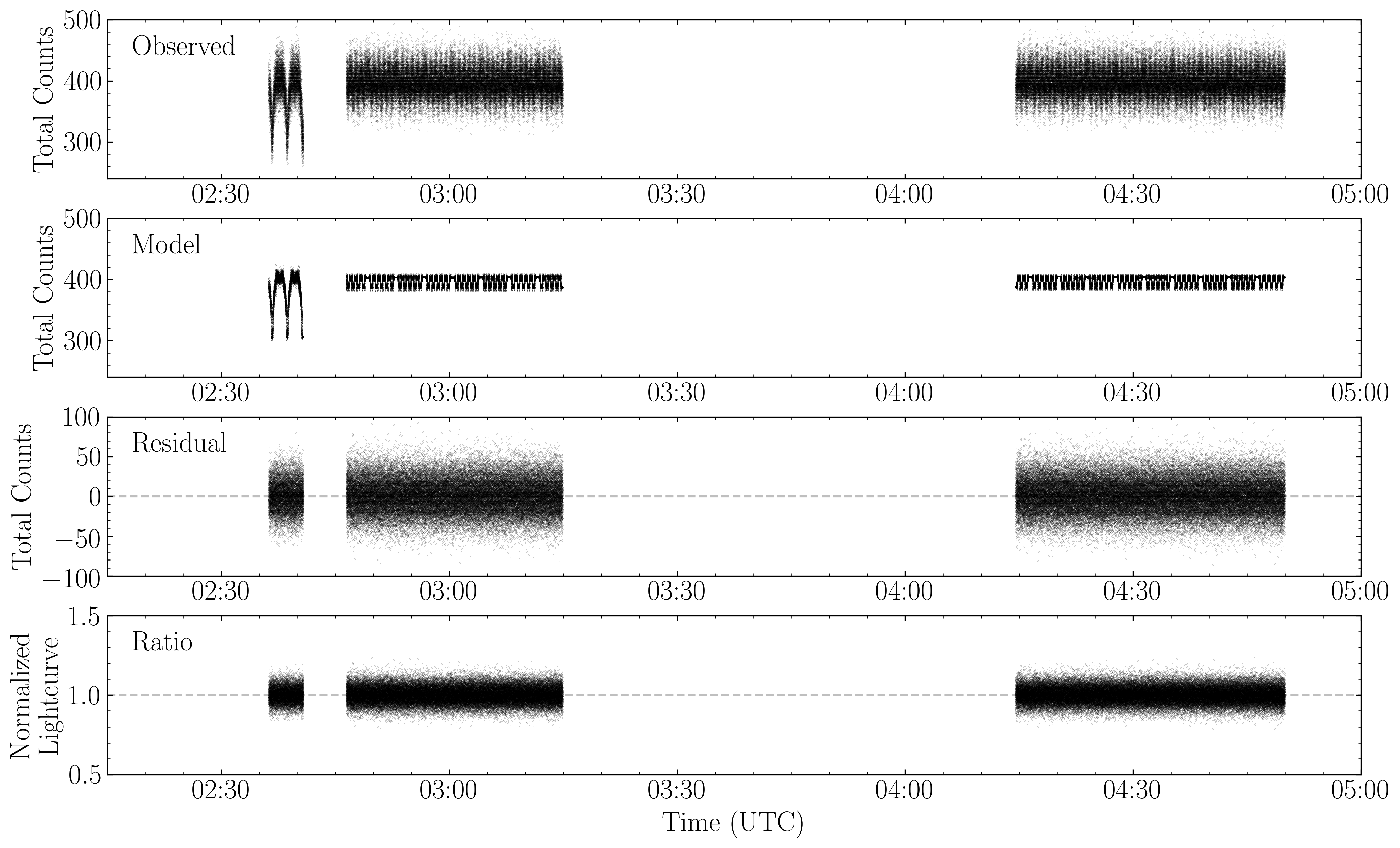}
\caption{Observed HST FGS counts during the occultation window, and the normalized lightcurve after correction for instrumental effects. Top panel: $C_{obs}$, the total counts measured by HST FGS. Second panel: $C_{inst}$, the modeled counts based on instrument effects for each segment of the observation. Third panel: $C_{obs}$-$C_{inst}$, the residual of model subtracted from observation. Bottom panel: $C_{obs}$/$C_{inst}$, the normalized lightcurve after correction.}\label{fig:SSmod}
\end{figure*}

This model allows for the removal of the periodic effects from the instrument scan itself before we search for transient, time-correlated signal from possible opacity along the line~of~sight to the star. The modeled counts ($C_{inst}$) are shown in Figure \ref{fig:SSmod}, along with the residuals ($C_{obs}$ - $C_{inst}$), and normalized lightcurve ($C_{obs}$/$C_{inst}$) derived from comparison with the observed counts. Poisson noise from the stellar source sets the signal to noise ratio (SNR), which can be estimated as $\sqrt{N}$ where $N$ is the number of observed counts during a given time interval. Therefore, we find that SNR$\sim$20 at the observed 40~Hz time resolution (or SNR$\sim$126 given a 1 second time interval).

\clearpage

\section{Upper Limits on Rings and Dust Opacity} \label{sec:limits}

\subsection{Numerical Modeling Method} \label{numerical}

While no signatures of opacity were immediately apparent in the photometry, we further adapted the forward model to determine upper limits on potential absorption by rings or dust in the nearby environment of MU69. We added a simple three parameter ring model for comparison: with $t_c$, the time at center of opacity during the observation; $\Delta t$, the duration of the occultation by a ring or other opacity feature; and $\tau$, the line~of~sight optical depth. We modeled the ring opacity as a boxcar function and tested durations of $\Delta t$ ranging from 0.25 to 3.5 seconds (approximately equivalent to ring widths of 5~km to 70~km). 

\citet{ortiz2017} and \citet{berard2017} modeled the rings around Haumea and Chariklo, respectively, with the assumption that they are co-planar with the equator of the main body. The rings also have some angle $B$, known as the ring-opening angle, or the angle between the ring plane and the line~of~sight vector of the observer. The pole orientation of MU69 is currently unknown due to the limited data available, so for simplicity and due to the lack of constraints from existing observations, we do not consider $B$ as a free parameter in this work. Instead, we assume a pole-on geometry ($B = 90^{\circ}$) for the rings. As a result, the line of sight optical depths reported in this work serve as upper limits to the normal optical depth $\tau_{N}$, defined as $\tau_{N} = \tau \sin{|B|}$. The normal optical depth is the optical depth that would be observed if the line of sight were perpendicular to the ring plane.

Based on this three parameter model, as a function of time the line~of~sight transmission is described as:

\begin{equation}\label{eq:Tmod}
T(t,t_c,\Delta t,\tau) = \begin{cases}
1 &\mbox{if } t < t_c - \frac{\Delta t}{2} \\
e^{-\tau} &\mbox{if } t_c - \frac{\Delta t}{2} \leq t \leq t_c + \frac{\Delta t}{2} \\
1 &\mbox{if } t > t_c + \frac{\Delta t}{2} \\
\end{cases}
\end{equation}

\noindent which leads to the modified version of Equation \ref{eq:SSENC2}:

\begin{equation}\label{eq:Cmod}
C_{model}(t,t_c,\Delta t,\tau) = T(t,t_c,\Delta t,\tau)\: \frac{\sum_{i=1}^n w_i(t)\: C_{obs}(t_i)/T(t_i,t_c,\Delta t,\tau)}{\sum_{i=1}^n w_i(t)}
\end{equation}

This step requires an additional factor of $T(t_i,t_c,\Delta t,\tau)$ within the summation in order to account for the secondary effect of opacity on the instrument correction itself. Without this factor, the effect of any dips in $C_{obs}$ due to opacity along the line of sight would bias the calculation of $C_{inst}$ and by extension $C_{model}$. Thus, the potential biasing effect of attenuation can be effectively negated by dividing $C_{obs}$ by the model transmission as shown here.

We demonstrate a few examples of potential ring opacity by using a synthetic lightcurve. The synthetic data, $C_{syn}$, was calculated as:

\begin{equation}\label{eq:synth1}
C_{syn}(t) = C_{inst}(t) + \mathcal{N}(0,\sigma^2)
\end{equation}

\noindent where $\mathcal{N}(0,\sigma^2)$ represents a random draw from a normal distribution, centered on zero, with the same standard deviation as the observed data. Adding opacity is done in a straightforward manner as:

\begin{equation}\label{eq:synth2}
C_{synmodel}(t,t_c,\Delta t,\tau) = T(t,t_c,\Delta t,\tau)\: C_{syn}(t)
\end{equation}

\begin{figure*}[ht!]
\plotone{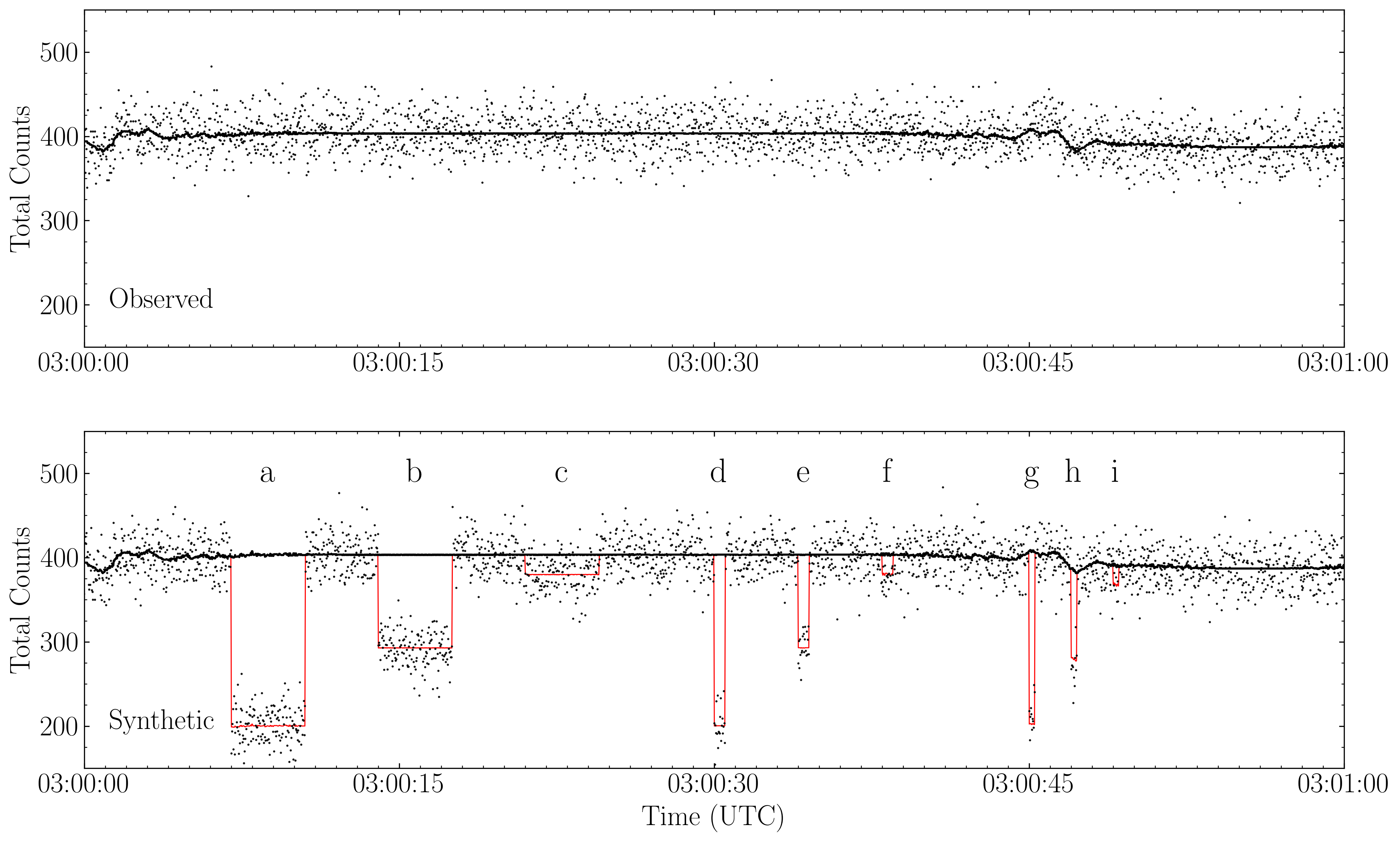}
\caption{Lightcurve for various putative ring width and opacity cases compared to observed data. Top panel: Observed data, $C_{obs}$ (black points), compared to $C_{inst}$ (black solid line). The apparent wiggles in the baseline count levels are due to changes in the instrument scan position, and are well matched by the correction from Equations \ref{eq:SSENC1} and \ref{eq:SSENC2}. Bottom panel: Synthetic data, $C_{synmodel}$ (black points), was generated from $C_{inst}$ (black solid line), with additional attenuation by parameterized ring or dust features as described in Equations \ref{eq:synth1} and \ref{eq:synth2}. The same synthetic data without added noise is also shown for comparison (solid red line). Parameters used for ring models a-i are given in Table \ref{tab:parameters}. \label{fig:synlightcurve}}
\end{figure*}

We generated several cases of synthetic ring absorption (models a-i), which represent the various combinations of ring model parameters shown in Figure \ref{fig:synlightcurve}. We also compare these model parameter values to previously detected ring systems in Table \ref{tab:parameters}.

In order to determine upper limits on opacity, we analyzed the parameterized model $C_{model}(t,t_c,\Delta t,\tau)$ and the observed $C_{obs}(t)$ within a Bayesian statistical framework. For computational efficiency, we calculated the posterior log likelihood of the model with a few selected values of $\Delta t$ over all reasonable values of $t_c$ and $\tau$. The log likelihoods of the null case (no opacity) and the model with a given set of ring parameters were derived from the following set of equations:

\begin{equation}\label{eq:logL1}
y(t) = C_{obs}(t)
\end{equation}

\begin{equation}\label{eq:logL2}
\sigma (t) = \sqrt{y(t)}
\end{equation}

\noindent If $\tau = 0$, as in the null case, then the following is true:

\begin{equation}\label{eq:logL3a}
C_{model}(\tau = 0) = C_{inst}(t)
\end{equation}

\begin{equation}\label{eq:logL4a}
\text{logL}(\tau = 0) = -\frac{1}{2}\sum_{i=1}^n \frac{(C_{obs}(t_i)-C_{inst}(t_i))^2}{\sigma(t_i) ^2}
\end{equation}

\noindent otherwise for the more general case:

\begin{equation}\label{eq:logL4b}
\text{logL}(t_c,\Delta t,\tau) = -\frac{1}{2}\sum_{i=1}^n \frac{(C_{obs}(t_i)-C_{model}(t_i,t_c,\Delta t,\tau))^2}{\sigma(t_i) ^2}
\end{equation}

\begin{center}
\begin{table*}[ht] 
\caption{Model Ring Parameter Space}
\label{tab:parameters}
\centering
\scalebox{.8}{
{\renewcommand{\arraystretch}{1.25}
\begin{tabular}{ccccc} 
\hline\hline
{Ring Model} & {w (km)}  & {$\Delta t$ (s)} & {$\tau_{N}$} & {Source}  \\	   				
\hline
Haumea				& 70.0	& -		& 0.70	& \citet{ortiz2017} \\
Chariklo CR1		& 7.0	& -		& 0.32	& \citet{braga2014} \\
Chariklo CR2		& 3.5	& -		& 0.06	& \citet{braga2014} \\
Saturn G Ring		& 10,000& -		& $10^{-6}$ & \citet{hedman2007} \\
a					& 70.0	& 3.5	& 0.70	& \\
b					& 70.0	& 3.5	& 0.32	& \\
c					& 70.0	& 3.5	& 0.06	& \\
d					& 10.0	& 0.5	& 0.70	& \\
e					& 10.0	& 0.5	& 0.32	& \\
f					& 10.0	& 0.5	& 0.06	& \\
g					& 5.0	& 0.25	& 0.70	& \\
h					& 5.0	& 0.25	& 0.32	& \\
i					& 5.0	& 0.25	& 0.06	& \\
\hline\hline
\end{tabular}}}
\bigskip
\end{table*}
\end{center}

\begin{figure*}[ht!]
\plotone{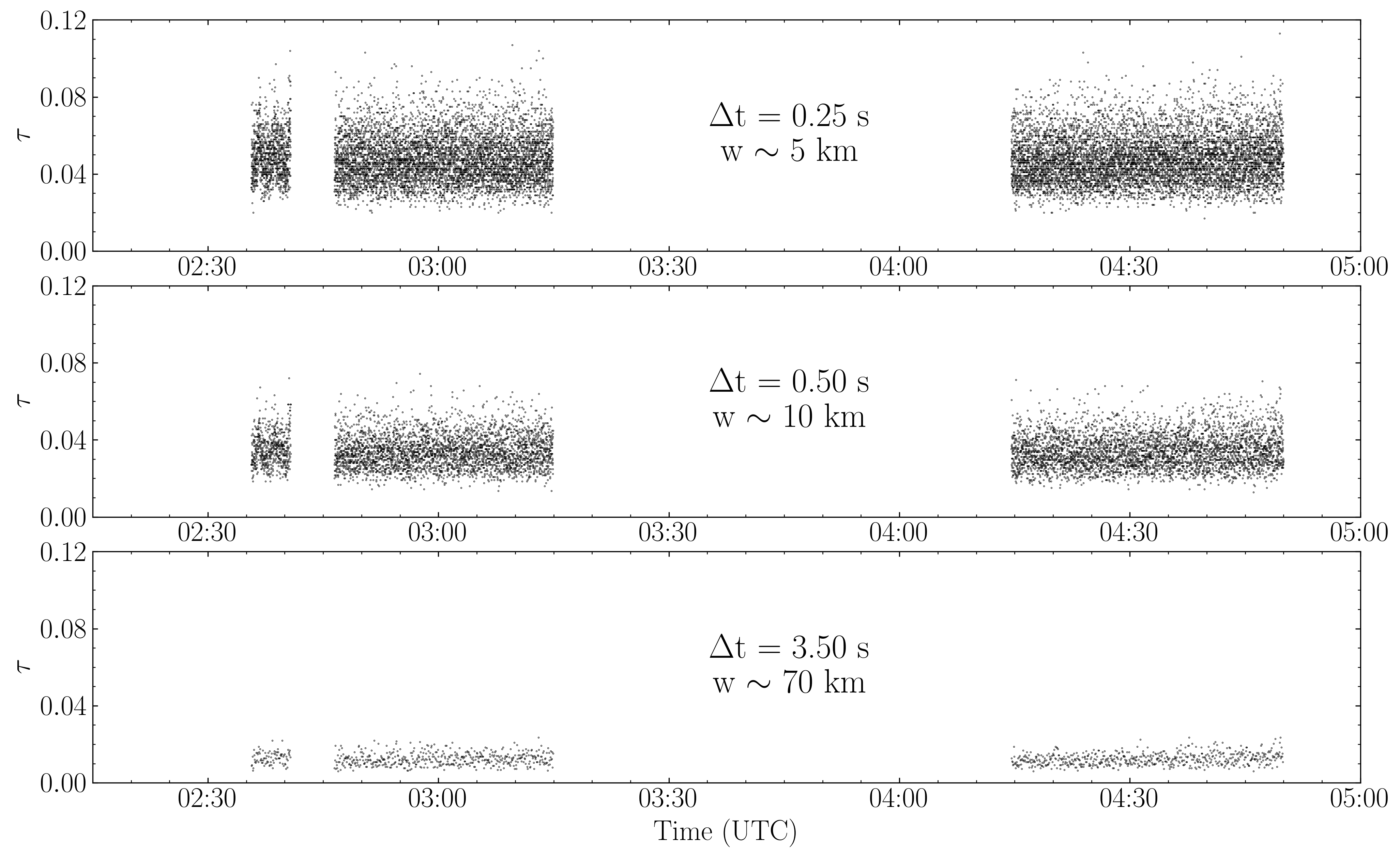}
\caption{3-sigma (99.7\% cumulative probability) upper limits on $\tau$ during the observation given a model value of $\Delta$t (or w, the equivalent ring width). The model ring widths shown were chosen to roughly correspond to the observed ring parameters of Haumea and Chariklo (see Table \ref{tab:parameters}). \label{fig:tau-upper}}
\end{figure*}

For a given value of $t_c$ (selected from the full range of times during the observation) and $\Delta t$ (selected from 0.25, 0.5, or 3.5 seconds), we calculated the posterior log likelihood of the opacity model for values of $\tau$ ranging from 0.001 to 0.1 in increments of 0.001. From these posterior likelihood distributions of $\tau$, we then estimated the 3-sigma (99.7\% cumulative probability) upper limit on opacity. This final product is shown in Figure \ref{fig:tau-upper} as a function of $t_c$ (time in UTC) and for $\Delta t$ of 0.25, 0.5, or 3.5 seconds. These values of $\Delta t$ are equivalent to distance scales of 5~km, 10~km, and 70~km given an approximate sky-projected velocity of $\sim$20 km/s during the occultation.

\subsection{Analytical Method} \label{analytical}
We verified the upper limits on the ring optical depth derived in the previous section using an analytic approach. We looked for statistically significant outliers in the dataset at various spatial resolutions, following a similar technique used by \citet{throop2015} to search for rings around Pluto during the \textit{New~Horizons} flyby. 

\begin{figure*}[ht!]
\plotone{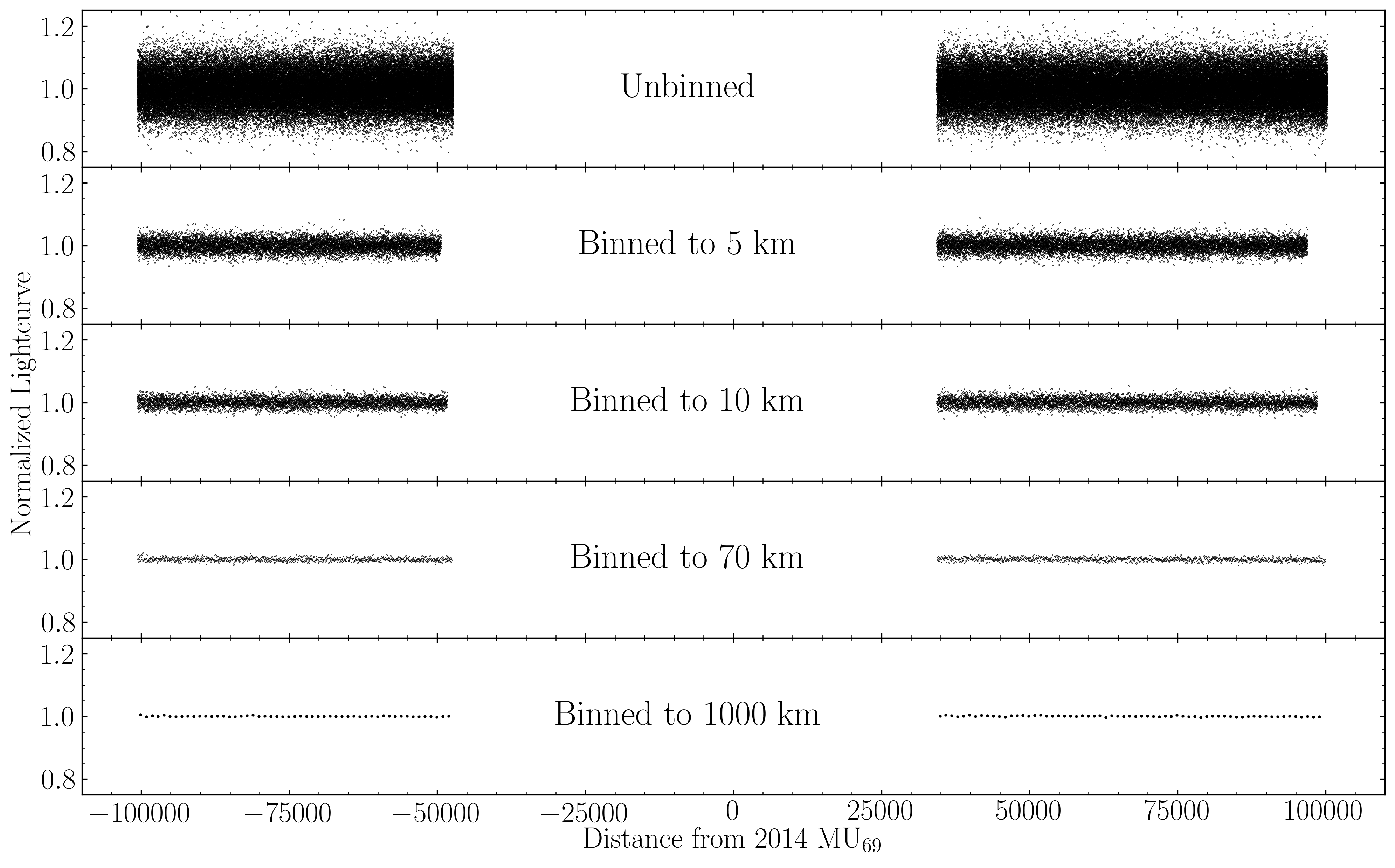}
\caption{Normalized lightcurves binned to different spatial resolutions. No statistically significant signatures of opacity were found in any of the binned datasets. \label{fig:tau-analytic}}
\end{figure*}

We applied this approach only to the final two focused scans, excluding the initial data that was collected to rule out the presence of a binary companion. For these N = 153,330 data points, we would expect that about 400 data points (or 0.26\%) would fall above or below the 3-sigma range and find 428 such points in the data. In Figure~\ref{fig:tau-analytic}, we bin the data to spatial resolutions comparable to the rings discovered around other small bodies and also explore the case of an extremely extended (1000~km wide) ring. For the case of the 1000 km, 70 km, 10 km, and 5 km binned data sets, we would expect to see 0.2, 4.5, 29.3, and 53.6 data points fall outside of the 3-sigma level, respectively. We find 0, 4, 37, and 48 points for each of those cases, further indicating that there are not likely any statistically significant outliers that would be suggestive of a ring or debris. 

We use this additional information to place upper limits on the optical depth of a putative ring around MU69. The equation for optical depth is defined as $I = I_0\, e^{-\tau}$, where $I$ and $I_0$ are the occulted and unocculted stellar signal, respectively. We used this relation to solve for the optical depth of a ring that would cause a 3-sigma detection in each dataset. These results are listed in Table \ref{tab:analytic} and are consistent with the upper limits on optical depth derived from the numerical model analysis discussed in the previous section.

\begin{center}
\begin{table*}[ht] 
\caption{Upper Limits on $\tau$: Analytic Results}
\label{tab:analytic}
\centering
\scalebox{.8}{
{\renewcommand{\arraystretch}{1.5}
\begin{tabular}{c|ccccc}
\hline\hline
Bin Width & No Binning & 5 km & 10 km & 70 km & 1000 km \\
{$\tau$ (3-sigma limit)} & 0.146 & 0.057 & 0.042 & 0.017 & 0.005 \\
\hline\hline
\end{tabular}}}
\bigskip
\end{table*}
\end{center}

\section{Discussion} \label{sec:discussion}

The HST FGS observation of the July 17, 2017 stellar occultation was successful and provided upper limits on sources of opacity within the Hill sphere of MU69. While HST was unable to observe the target star during the mid-point of the occultation, the results still probed down to radii of about 20,000~km from the main body. At these distances, the data rule out significantly opaque sources at effective ring widths of $\sim$5~km or more. 

These results provide additional context for the summer 2017 campaign of MU69 stellar occultation measurements made by ground- and air-based observatories as reported in \citet{buie2018} and elsewhere. These collective upper limits on sources of opacity in the MU69 system will be invaluable as planning continues for the January 1, 2019 flyby of MU69 by \textit{New Horizons}.

\acknowledgments

Support for Program number HST-GO-15003.001-A was provided by NASA through a grant from the Space Telescope Science Institute, which is operated by the Association of Universities for Research in Astronomy, Incorporated, under NASA contract NAS5-26555. Additional support for this effort was provided through the NASA New Horizons project.

\vspace{5mm}
\facility{HST(FGS)}

\end{document}